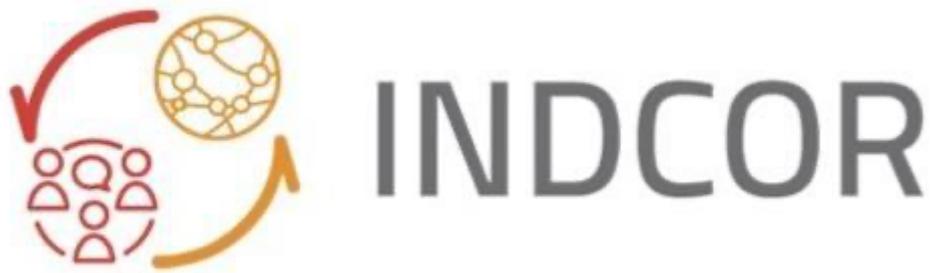

INDCOR White Paper 0

Interactive Digital Narratives (IDNs) –

A Solution to the Challenge of Representing Complex Issues

Koenitz, H., Barbara, J., Holloway-Attaway, L., Nack, F., Eladhari, M. P., Bakk, A.

**INDCOR White Paper 0: IDNs representing complex issues**

**Executive Overview**

Citizens everywhere have the right to be well-informed. Yet, with the high complexity of many contemporary issues, such as global warming and migration, our means of information need to mutually adapt. Narrative has always been at the core of information exchange – regardless of whether our ancestors sat around a fire and exchanged stories, or whether we read an article in a newspaper, or watched a TV news broadcast. Yet, the narrative formats of the newspaper article, the news broadcast, the documentary, and the textbook are severely limited when it comes to representing highly complex topics which may include several competing – and sometimes equally valid – perspectives. Such complexity contributes to a high level of uncertainty due to a multitude of factors affecting an outcome. Fortunately, with Interactive Digital Narrative (IDN), there is a novel media format which can address these challenges. IDNs can present several different perspectives in the same work, and give audiences the ability to explore them at will through decision-making. After experiencing the consequences of their decisions, the audience can replay to revisit and change these decisions in order to consider their alternatives. IDN works enable deep personalization and the inclusion of live data. These capabilities make IDN a 21$^{st}$ century democratic medium, empowering citizens through the understanding of complex issues. In this white paper, we discuss the challenge of representing complexity, describe the advantages offered by IDNs, and point out opportunities and strategies for deployment.

Four additional white papers, linked from this one, offer more detailed insights into **design**, **conceptions**, **evaluation,** and **societal impact**.

**INDCOR White Paper 0: IDNs representing complex issues**

Introduction

Today, we are faced with numerous highly complex topics such as global warming, migration, and pandemics. The complexity of those topics is based on their systemic nature – the interplay of elements in and between various systems (both natural, such as ecological or meteorological, as well as human generated systems, such as economic, juridical, social, and technical). The question is how can we understand these issues, discuss them with others, and make well-informed decisions which are prosocial, sustainable, personally gratifying and healthy?

Humans have used various ways of storytelling throughout history to represent, communicate, discuss and transfer knowledge about the world ever since we sat around fires exchanging stories. In principle, these stories or 'narratives' can still serve the same role for increasingly complex issues in 21$^{st}$ century digital culture. However, a particular and important challenge is that established forms of narrative media (i.e., newspaper articles, TV features, documentaries, movies, and books) are fixed due to the nature of the medium (print, video, and audio). And as such, they are limited in their ability to represent highly complex topics due to the latter's dynamic nature.

For audiences to gain a systemic understanding, it is not enough to tell them about underlying relationships and results impacted by many factors. Rather, audiences must be able to experience such complex systems themselves. Crucially, IDNs possess the qualities necessary to enable dynamic, self-guided and personalized experiences and thus facilitate systemic understanding, something traditional media is unable to provide.

> ***To gain a systemic understanding, it is not enough to tell audiences about complexity – they need to be able to experience the complex systems themselves.***

IDN combines computational dynamic systems and narrative in a novel compelling media format. IDNs react to decisions made by the audience to create a customized narrative experience. Furthermore, audiences can experience many different views through replay and learn how a series of decisions affect each other and lead to a specific outcome. They can even add their own perspectives and discuss their insights with others. Taken together, these characteristics facilitate a deep and systemic understanding of complex issues and make IDNs a fitting media format for the 21$^{st}$ century.

**INDCOR White Paper 0: IDNs representing complex issues**

**The Challenge of Complexity in Postmodern Societies**

In democratic systems, non-fiction communication plays a crucial role in helping citizens understand the world around them and make informed decisions in their private lives and as political actors. Indeed, the social theorist Jürgen Habermas understood citizens as informed newspaper readers (Habermas et al., 1974). However, we no longer live in a period where newspapers are the dominant medium for news distribution.[1] Instead, social media, online information sites, and TV channels compete for the audiences' attention. Similarly, with the increasing complexity of many contemporary issues (e.g., economic globalization, global warming, migration, pandemics, and terrorism), political decision-making has equally become more complex. In addition, there is a foundational conflict between the ideals of objectivity and truth on the one hand, and complex phenomena which lack such stable categories, and are characterized as 'multi-truth,' 'multi-perspective' and 'emergent', on the other. We can see the French Philosopher Jean-Francois Lyotard's insight of postmodernity as a "crisis of representation" (1984), as a warning about the limits of modernist, fixed representations. An example is the challenge of representing the issues of aging populations and their varied effects on economic and social development in fields like health care, urban development, and the available workforce. Certainly, truth and objective reporting remain important values, but when there are no longer a limited number of ideologically delimited 'truths', but instead a complex space of possible solutions, then we need new ways to report, inform, and educate.

The new standard way to approach such issues should be "systemic thinking," as Rejeski et al. remind us in a research report for the Wilson center (2015). Systemic thinking is the ability to understand the systemic dynamic nature of many contemporary phenomena. This perspective is crucial for understanding both the challenges of, and the possible approaches towards, complex issues. However, so far, systemic thinking has only taken a limited hold in the representation of politics, both on the side of political actors and of journalists. One aspect of this crisis is a disconnect between actual political decision-making and an often outdated, simplistic ideological frame narrative, in which the same political actors are caught. To make matters worse, in the absence of successful strategies for representations of complexity, purveyors of simplistic narratives feed many people's longings for the 'easy,' modernist narratives of time past, as indicated by the increasing votes for populist parties in many western democracies, who often perpetuate these simple narratives as forms of persuasion and influence.

---

[1] See for example the Reuters 2022 Digital News Report:
https://reutersinstitute.politics.ox.ac.uk/digital-news-report/2022/dnr-executive-summary

*The structural problem of traditional media*

Contemporary media reporting in newspapers, TV newscasts, documentary and other established media formats are frequently criticized for their shallowness and undue over-simplification when representing complex issues. Political actors – especially populist ones – portray these issues as evidence for hidden political agendas and opaque, partisan side-taking. Indeed, cases of undue bias, fraudulent reporting and lack of competence certainly exist. Yet, the underlying problem and root cause behind inadequate representations of complex issues is mostly misunderstood and consequently wrongly attributed. Particular media producers, media organizations, or even the topics themselves are not the primary problem, but rather a structural problem exists when it comes to representing such complex issues, a problem that so far has often been overlooked. Static forms of representation such as newspaper articles and TV reports, by their very nature, reduce and simplify complex phenomena, to non-dynamic, fixed representations, that eliminate the systemic and dynamic aspects, and thus create misrepresentations.

> ***Static forms of representation, such as newspaper articles and TV reports, by their very nature, reduce and simplify complex phenomena, to non-dynamic, fixed representations that eliminate the systemic and dynamic aspects and thus create misrepresentations.***

Fixed media forms may provide partial depictions of crucial issues and social phenomena, but struggle to provide complete and complex representations, as they cannot provide a truly dynamic and individual experience. For example, when focusing on the issue of climate change, a newspaper article might describe rising seawater levels, and a TV program might show how human-made emissions affect weather patterns, but they cannot provide a dynamic and personalized experience where audiences can actually see their own carbon footprint, or compare different arguments with scientific evidence, or in a simulation, even try different measures to reduce the effects of climate change. These limitations are especially problematic in democratic societies, which depend on well-informed citizens for public discourse and decision-making. To facilitate a well-informed citizenry of the 21st century, new media formats such as IDNs are necessary for the representation of the highly complex topics of our time.

**IDNs Representing Complexity**

In general, IDNs are narrative experiences that can be changed by an audience, and which are created for the digital medium. What the audience can change (for example, how the narrative progresses, which perspective to follow, and what ending is reached) varies as does the particular form. Popular IDN forms include narrative-focused video games, interactive documentaries (i-docs), journalistic interactives and narrative-driven eXtended (Virtual/Augmented/Mixed) Reality (XR/VR/AR/MR) experiences.

The following range of applications demonstrate the potential for IDNs as representations of complexity on the societal, as well as personal, levels. For example, when it comes to political decision-making, the challenge often lies in understanding the complex, immediate, and long-term effects of policies. A good entry point to this topic are the political decision-making IDNs from NewsGamer[2]. *Collapsus* is another example in this regard, focused on future energy security from Amsterdam-based media production house Submarine Channel[3]. Similarly, *Survive the Century*[4] is concerned with climate change and puts a player in the role of a decision-maker (i.e., a news editor) whose choices determine if humanity can be saved in the face of a growing climate crisis. It was produced by an international team from the US and South Africa. Narrative-focused games such as these can also provide players with personal experiences from specific political/historical situations, for example, being a human rights monitor in *Starlight Stadium* (OSCE ODIHR, 2023) being part of a resistance group in WW2 as in *Attentat 1942* (Charles Games, 2017) or as a judge during the French revolution as in *We, The Revolution* (Polyslash, 2019). The complex issue of Somali piracy is explored in the interactive documentary The *Last Hijack Interactive* (Duijn et al., 2014) where the interactor is given the opportunity to explore different perspectives surrounding a particular hijacking incident, including interviews with the victims, perpetrators, and negotiators. The economy of the illegal drug trade in the Netherlands is the subject of *The Industry* (Duijn, 2018), Here, interactors can explore what drug-related activities happened around a particular Dutch zip code, learn about the shallow organization structure, and follow audio interviews with different players in the industry. Complexity also features in many narrative-focused video games. For example, *Papers, Please* (Pope, 2013) explores the conditions of life in totalitarian regimes. Here, the interactor plays the role of a border control officer, under pressure of constantly changing instructions and retributions in case of errors, responsible for his family's survival and also facing the fact that some visitors are actual terrorists who will commit murders if allowed through. Likewise, the video game simulation *The Sims* (Wright, 2000) brings to the fore aspects of the complexity of life in late capitalism, including earning a living, reaching life objectives, finding a partner, and raising a family. The interactive drama *Façade* (Mateas & Stern, 2003) explores the complexity of a relationship at the breaking point, while *PromWeek* (McCoy et al., 2011) explores the complex social dynamics surrounding a high school prom.

---

[2] www.newsgamer.com
[3] https://submarinechannel.com/game/collapsus-energy-risk-conspiracy/
[4] https://survivethecentury.net

> **Conceptual Understanding of IDN: the SPP Model**
>
> IDN has three components: system, process, product. The SPP model (Koenitz, 2010, 2023) understands a dynamic *system* as the central element and describes the interactive *process* which results in instantiated narrative *products*. The latter are either objective products (recordings and play traces) or subjective products (retellings (Eladhari, 2018)). The contents of the system form a *protostory* which includes fixed *assets* such as graphics and 3d models, *UI* (user interface) and *procedural components*, such as rules and generative systems, brought together in the *narrative design*. (Figure 1).
>
> 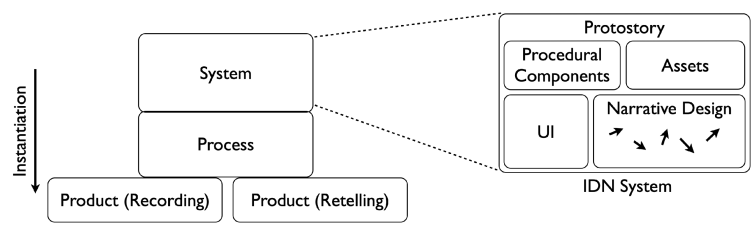
>
> Figure 1. High-level view of the SPP model (left) and protostory elements (right).

**INDCOR White Paper 0: IDNs representing complex issues**

**The Problem Space of Complexity and IDN**

The case for IDN as a media format for the 21st century is compelling. However, today, IDNs are not yet part of the mainstream media. The following statements describe the current situation:

- There are few expert interactive narrative designers, especially outside of video game design, and the design knowledge that exists is limited and isolated.
- Due to the so far limited application cases, the literacy of interactive forms for informational and educational purposes is limited amongst mainstream audiences.
- IDNs (and other interactive forms) are considered as experiments by mainstream media producers, which prevents audiences from familiarizing themselves with this new format.
- Interactive forms such as IDNs are taken as peripheral by many established media production companies, as some have no in-house expertise or production capacity.
- No standardized production workflow for IDNs exists, in contrast to newspaper, textbook, or TV news production.
- Funding options for interactive productions are limited in many countries, in contrast to specific funds, tax breaks and other incentives which exist for film production and other more established formats.
- More research is necessary to understand the most effective application of IDN to represent complex phenomena in an understandable form without trivializing it.

To better understand the challenges encountered by IDNs as representations of complexity, it is important to realize that this novel format exists in an ecology of established approaches, practices, and products: a problem space (see Fig. 2) that connects professional training, education, IDN production and funding, and research.

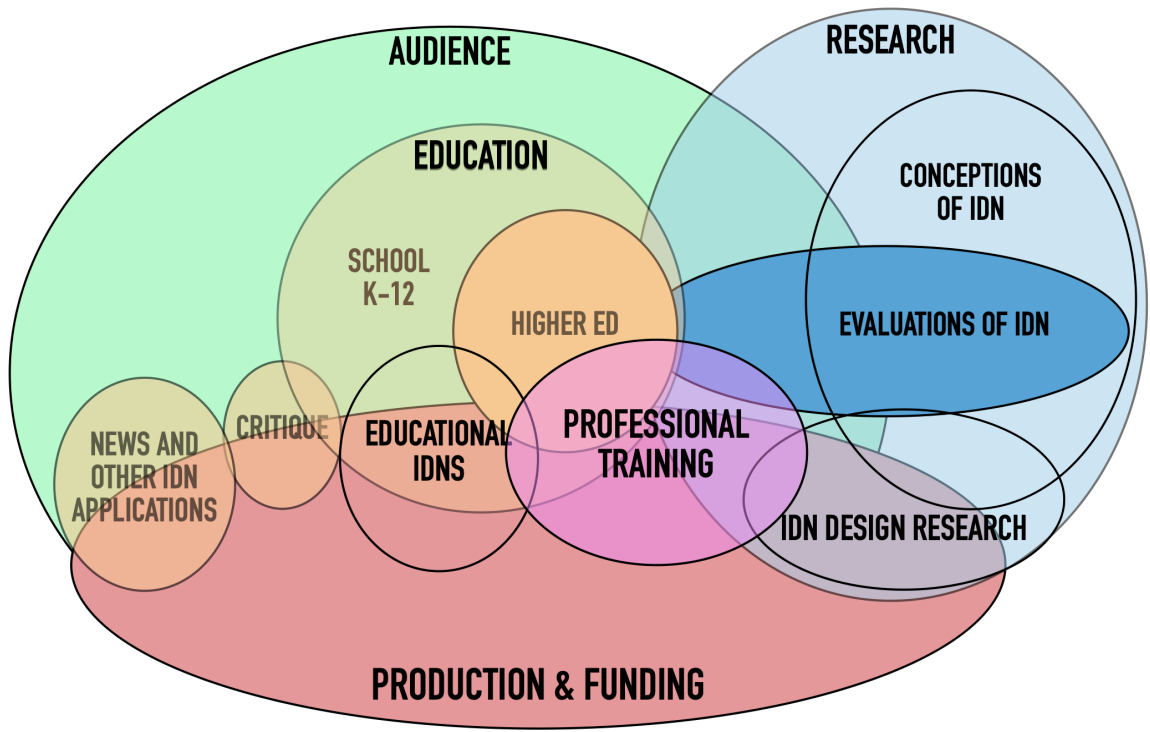

Fig. 2. A visualization of the problem space of IDN and complexity

**Turning IDNs into a widely-used format**

To address these challenges, a number of interconnected measures are necessary, including professional training of creators, education in IDN literacy across all age groups[5], changes to media production workflows[6], and funding structures adapted to accommodate IDNs. In addition, researchers need to further improve methods to evaluate and verify the effectiveness of IDN applications.

*Professional Training*

For many media creators, working on IDNs means a paradigm shift, as they will no longer produce static products, such as movies or newspaper articles for viewers and readers, but dynamic works, which are made for an audience that has an active role as participants in shaping the experience. These creators are no longer the media producers of old, but architects of novel, unique dynamic systems. This means to (re-)train existing media creators such as journalists as IDN system builders, but also to start offering IDN-focused programs in higher education.

---

[5] In line with the European Media and Audiovisual Action Plan's focus on audiovisual entertainment sector specifically in its aim to Enable and Empower by enhancing media literacy. https://digital-strategy.ec.europa.eu/en/policies/media-and-audiovisual-action-plan

[6] In line with the News Initiative to unleash innovation by providing innovative applications and technologies in the media domain. https://digital-strategy.ec.europa.eu/en/policies/news-initiative

Such training and programs need to cover the novel ethical dimension inherent in the multifaceted representation of complex issues. The development of a code of conduct – similar to the ones for journalism – is, thus, an important topic for further investigation. A first step in this direction has been taken by the introduction of an ethics framework for IDN authoring (Koenitz et al., 2023).

*Education*

There is also a need to rethink our educational systems. Both the source of learning (in the form of printed or video materials) and the standard reporting format (essays, theses) are fixed narratives. These methods are particularly ill-equipped to prepare pupils for a complex reality. It is therefore crucial for policy-makers to take note of the benefits of IDN in education and consider integrating IDNs into school curricula. The aim should be to achieve IDN literacy across all educational levels, while also providing teachers with IDN-related skillsets.

Finally, the education of audiences about IDNs is an area in need of attention, as is a focus on critique of IDNs (in line with film criticism) in high-profile outlets, such as in broadsheet newspapers and TV channels, to further educate adult audiences about IDNs.

*Production and Funding*

Currently, IDNs are still predominantly one-off projects and, while regular production capacities have been established (e.g., by big US newspapers such as the *New York Times*), there is still a documented tendency[7] to use IDNs as a means to drive subscribers to the "regular content" of newspapers' websites. For IDNs to become a standard media format, a number of measures are necessary, including the integration of IDN into established production workflows, the already mentioned (re)training of media creators and changes in the funding structures. Currently, many national systems of funding are only available for traditional media formats, such as documentaries, but not for interactive content. Possible solutions to this problem include the extension of existing public broadcasting funding models to cover IDNs, and also the adoption of business models similar to the ones used by commercial video games and newspapers.

---

[7] https://edspace.american.edu/jolt/

*Research*

The interdisciplinary endeavor of the EU COST action INDCOR is addressing these issues with the aim of providing the fundamental material for IDNs to become "a means to address complexity as a societal challenge."[8] INDCOR addresses the above aspects in different work groups and their specialized knowledge is available in the form of four separate white papers:

- IDN Design (WG1)

  The white paper functions as an orienting map in the field of IDN design. It highlights the issues of complexity, of its representations, and how Interactive Digital Narratives could be designed to address them. A wide range of topics related to IDN design is presented, including IDN workflow and methodology, IDN authoring tools, and IDN applications.
  Link: https://arxiv.org/abs/2305.01925

- Conceptions of Interactivity (WG2)

  This text is concerned with the complexities of authoring, delivering, and processing dynamic interactive contents from the perspectives of both creators and audiences. It examines the range of user-content interactions associated with IDN and discusses how various interaction modes can affect both the representation of interactive narrative content and related to this the users' comprehension.
  Link: https://arxiv.org/abs/2306.10547

---

[8] Quoted from INDCOR project homepage, https://indcor.eu

- Evaluation of IDN works (WG3)

    This white paper focuses on evaluation, presenting empirical methods employed by INDCOR members. It considers the impact of IDNs on audiences through evaluation, providing the means to gain a thorough understanding of the short term and long-term effects of engagement with complexity IDNs. Link: http://arxiv.org/abs/2306.09817

- Societal Impact (WG4)

    This white paper focuses on IDNs as powerful tools to address complex societal challenges, including racism, coloniality, feminism, war, cultural heritage and media disinformation. Six case studies of IDNs are presented to demonstrate how IDNs can engage users and allow them to actively participate in the complex social challenges they address. The white paper is offered as inspiration for journalists, digital content creators, game designers, developers educators and any others who might want to learn how the design and development of IDNs can support and engage audiences in complex societal issues to support social change.

    Link: https://arxiv.org/abs/2306.09831

**Opportunities**

In this section, we will briefly identify opportunities presented by the novel media format of IDN. We are at an important juncture in the digital 21$^{st}$ century, where social networks enable anyone to create and circulate content, enabling both the circulation of useful information and dis/misinformation ("fake news") which has the power to negatively impact societies. As we have argued, IDNs can address these challenges in strategic ways and consequently offer unique opportunities for the further development of public institutions and private enterprises in democratic societies.

IDNs in formal education can help students develop capacities that support systemic thinking. By critically engaging with IDNs and by creating IDNs themselves they can acquire the skills necessary to act in real world scenarios defined by complexity[9]. Additionally, we can use IDNs to introduce critical social and historical issues (e.g., war, politics, climate change, and racism), immersing students in the complexities and multi-perspectives that inform and uphold different value systems and power structures. New curricula need to be written to make use of the opportunity of IDN and train students to see the world and its narratives (i.e., histories, politics, and identity-making) as complex, multi-tiered, creating a novel understanding that is vital to 21$^{st}$ century democracy.

Another opportunity exists in the representation of cultural and natural heritage. Museums have often been criticized for being places that support and 'put on display' problematic artifacts representing oppressive interests. These can be subverted by presenting IDNs that allow the visitor to take the perspective of the oppressed and thus make the audience aware of issues that are glossed over in traditional univocal narratives. Audiences can gain a new understanding through strategic decision-making and the experience of competing value systems.

Similarly, we can challenge the way news is reported and circulated[10]. We can create IDNs which allow citizens to engage with big social issues, such as political decision-making and climate change, allow them to see several perspectives, and also to experience the mechanics behind news reporting and decision-making, for example by experiencing how a news story is shaped by journalists and received by an audience. By enabling interactive content for news stories, we can shape the reporting so it includes multiple voices but does not descend into toxic 'trolling,' and thus discover new ways to engage citizens.

The realization of these opportunities depends also upon policy-makers. In this regard IDNs have a dual role – as the subject of policy-making, but also as an opportunity to implement novel ways to facilitate policy-making itself. To support IDNs in education systems, not only new curricula, but new pedagogical policies and forms of teacher training need to be created. In addition, research around the educational use of IDNs needs to be increased and its results widely disseminated. To support policy-makers, IDNs can be created that show different perspectives on policy decisions, but also possible consequences.

---

[9] In line with the EU's News Initiative that aims to nurture democracy by 'help[ing] citizens of all ages navigate the modern news environment, take informed decisions and be better equipped to face the threats of disinformation' https://digital-strategy.ec.europa.eu/en/policies/news-initiative

[10] IDNs can serve as a new form of production and distribution of data-driven news as supported by the EU's Multimedia actions https://digital-strategy.ec.europa.eu/en/policies/multimedia-actions

Given these opportunities, the future market for IDN training and production is considerable and could easily reach billions of euros, mirroring the existing markets for textbooks and traditional news production as well as the related professional training. IDN provides us with the opportunity to create new businesses (e.g., companies specializing in the creation of interactive museum exhibits), but also entirely new business models in the creation of media content for public consumption that represents social complexity to serve as a foundation for discourse and decision-making in 21$^{st}$ century democratic systems. For example, an IDN could represent a proposed infrastructure measure and enable both citizens and policy-makers to directly engage with it and potentially change it.

**Conclusion**

In this white paper, we have made the case for interactive digital narratives (IDNs) as representations of complexity in different contexts. This novel media format can represent the many complex issues societies and individuals are facing in the 21$^{st}$ century. Crucially, IDN facilitates a deep, systemic understanding of complex problems such as migration, global warming, pandemics, energy transition, and personal crises by offering self-directed experiences of multiple perspectives, decision-making and resulting consequences as well as replay to explore different perspectives and choices. 21$^{st}$ century audiences have become accustomed to interactive forms and are, in principle, ready for novel interactive forms, but there is a lack of IDN literacy which can be addressed by consistent exposure, high quality production, and advocacy in high-profile settings such as broadsheet newspaper critiques. To fully capitalize on this potential and make IDN a standard format for reporting and education, a number of measures are necessary, in particular in the practice of media production and financing, as well as in the training of journalists and teachers. IDN production is also an opportunity for future-oriented media and cultural industry development and innovation. Higher education and academic research have an important role in the IDN ecosystem by providing training for next generation creators and further improving the understanding of design, applications and effects. The EU COST Action INDCOR is at the forefront of this task and its more than 200 members invite all interested parties to join us in this endeavor in tackling the complex issues of the 21st century. We believe IDNs represent a crucial, and timely opportunity for citizens as well as policy-makers to better understand the complexity of contemporary issues, improve the resilience of democratic institutions and facilitate positive social change.

**Acknowledgements**

The authors like to acknowledge the support of the EU COST Association (Cooperation in Science and Technology), through the COST Action 18230—Interactive Narrative Design for Complexity Representations (INDCOR). The authors thank the participants of INDCOR for many insightful discussions which have shaped this article.